\documentstyle[aps,prd,12pt]{revtex}

\begin{document}

\title{On Page's examples challenging the entropy bound}

\author{ Jacob D. Bekenstein\thanks{E--mail:
     bekenste@vms.huji.ac.il}}

\address{\it Racah Institute of Physics, The Hebrew University of
Jerusalem\\
Givat Ram, Jerusalem 91904, Israel}

\date{\today}
\tighten
\maketitle
\def\cu#1{\nabla\times{{\bf #1}}}
\def\cucu#1{\nabla\times\nabla\times{\bf #1}}
\def\di#1{\nabla\cdot{\bf #1}}
\def\ve#1{{\bf #1}}

\begin{abstract}
According to the entropy bound, the entropy of a complete physical
system can be universally bounded in terms of its circumscribing radius and
total gravitating energy.  Page's three recent candidates for counterexamples
to the bound are here clarified and refuted by stressing that the energies of
all essential parts of the system must be included in the energy the bound
speaks about.  Additionally, in response to an oft heard claim revived by
Page, I give a short argument showing why the entropy bound is obeyed at low
temperatures by a complete system.  Finally, I remark that Page's renewed
appeal to the venerable ``many species'' argument against the entropy bound
seems to be inconsistent with quantum field theory.  
\end{abstract}


\section{Introduction}
\label{intro}

In 1981 I proposed\cite{bek81} that the entropy of a complete physical system
in three spatial dimensions whose total mass-energy is $E$, and which fits in
a sphere of radius
$R$, is necessarily bounded from above,
\begin{equation}
S\leq 2\pi ER/\hbar c,
\label{bound}
\end{equation}
(henceforth I set $\hbar=c=1$).  The motivation comes from {\it gedanken\/}
experiments in which an entropy bearing object is deposited in a black hole
with a minimum of energy; a violation of the generalized second law seems to
occur unless the bound applies to the object\cite{bek81,bek94,bek99}.  A
variant {\it gedanken\/} experiment\cite{bek96} in which the object is
freely dropped into the hole again suggests that an entropy bound of the
above form must be valid.   A number of direct calculations of entropy of
systems with no black hole involved have supported the validity of the
bound for systems with negligible self-gravity\cite{bek94,bek84,review}. The
entropy bound~(\ref{bound}) is related to the holographic
bound\cite{bek96,bek00}, and is often a stronger restriction on the
entropy.  Because of the intimate relation of information to entropy, 
bound~(\ref{bound}) implies a fundamental restriction on the capacity of
information  storage and communication
systems\cite{bek84,review,bek81_b,bek88}.  This may become practically
important in the distant foreseeable future (the same cannot be said of the
vastly more lenient holographic information bound). 

Over the years Page has proposed a number of counterexamples to the entropy
bound\cite{page82,page1,page2,page3}.  In ref.\cite{bek82} I refuted the two
from ref.\cite{page82}, while in ref.\cite{bek83} I showed Page's proposed
substitute\cite{page82} for bound~(\ref{bound}) to be violated. In the present
paper I concentrate on Page's recent candidates for counterexamples, and show
that they do not constitute violations of the entropy bound as originally
stated. It is common for critics of bound~(\ref{bound}) to fail to include in
$E$ the energy of some part of the system, contrary to the
stipulations\cite{bek81,bek94,bek84,bek82} that the bound applies only to a
complete system---one that could be dropped whole into a black hole---and 
that $E$ in the bare inequality~(\ref{bound}) means the system's gravitating
energy (gravitating energy, of course, is immune to the well known
arbitrariness of the zero of energy).  Excluding part of the energy from $E$
empties the bound of meaning.  Thus, an harmonic oscillator with sufficiently
large mass (which makes the energy spacing of its levels small)  manifestly
violates the bound unless one includes the oscillator's mass in
$E$\cite{bek84}.   Yet in all his new examples, Page drops from $E$
the energy of some part of the system.

Page's first example\cite{page1,page2} is a self-interacting scalar field
with a symmetric double well potential, with the field vanishing at a certain
spherical boundary of radius $R$.  Page correctly points out that quantum
tunnelling between the wells splits the classically degenerate ground states
into a new ground state and an excited state a very small energy $\Delta E$
above the ground state.  Page hastily concludes that this violates
bound~(\ref{bound}) because the entropy of a mixed state built out of the two
can reach $\ln 2$, while $2\pi R\Delta E$ can be very small in natural units
($\hbar=c=1$).  As I show in Sec.~\ref{field}, Page's exclusion of the
classical energy of the configuration (much larger than $\Delta E$) is
unjustified because the last contributes to the field's gravitating energy. 
When $E$ includes it and the indispensable minimum wall energy,
bound~(\ref{bound}) is always obeyed.

In all of his new examples Page excludes contributions to $E$ from the
passive parts of the system.   He quotes  extensively from Schiffer's and my
paper\cite{SB} as license for this procedure.  The context is this.  In
the 1980's, based on the many cases studied numerically, we had toyed with
the idea that a strong form of bound~(\ref{bound}) might apply, whereby $S$
and $E$ in inequality~(\ref{bound}) refer solely to a collection of
{\it noninteracting\/} massless quanta, with the contributions to $E$ from
the cavity containing them ignored.  We conjectured a theorem to this
effect, which we proved for massless scalar quanta confined to a cavity of
arbitrary topology by Dirichlet boundary conditions\cite{review,SB}.  We also
sketched the proof of a generalization of that theorem for photons or
massless neutrinos; we considered only the wave equation, but did not go
into details about boundary conditions or constraint equations which are
very specific for electromagnetism or fermion fields.  Despite some efforts,
we  never completed these proofs.  So in the intervening decade I have used
bound~(\ref{bound}) in its original formulation\cite{bek94,bek96,bek00}. 
Page's new examples, which hark back to a version of the entropy bound
which did not become established generically, teach us little about the true
bound on entropy, and certainly do not constitute counterexamples to it.  
  
In his second example\cite{page1}, an onion-like arrangement of
a number of infinitely conducting concentric partitions separating
electromagnetic fields in different states of excitation, Page establishes
that some mixed states of the electromagnetic field disobey the strong bound
$S<2\pi(E-E_{\rm vac})R$, where $E-E_{\rm vac}$ refers only to the
electromagnetic energy above the vacuum,  and
$R$ to the radius of the outermost partition.  But this is {\it  not\/} a
violation of Schiffer-Bekenstein theorem\cite{review,SB}; the
theorem (strictly proved only for scalar quanta) only claims that the
`strong entropy bound' is obeyed by a {\it noninteracting\/} field in a
cavity, whereas in Page's onion the electromagnetic field {\it interacts\/}
throughout the volume of the sphere of radius $R$ with certain massive
charged fields describing the conducting material of the partitions.  My
semiclassical discussion of the conducting partitions  in Sec.~\ref{onion}
shows that even without specification of the  nature of the  charge
carriers, the complete system does obey the  original entropy
bound~(\ref{bound}).
   
Page's third example\cite{page2} is a closed loop of coaxial cable of total
length $L$ coiled up inside radius $R$, with $R\ll L$, which confines an
electromagnetic field.  Page gives hand waving arguments that the lowest
travelling frequencies are $2\pi/L$.  Citing again the electromagnetic
version of the Schiffer-Bekenstein theorem for license, Page estimates $E$ for
the cable containing a mixed state based on the two lowest lying modes to be
of order $1/L$.  Since the state can have entropy of order unity, Page argues
that the entropy bound is violated since $RE\ll 1$.  If Page is correct about
the low frequencies, then the Schiffer-Bekenstein theorem cannot be proved for
electromagnetic fields confined to a waveguide with a not simply connected
crossection.  At any rate, the complete system includes the charged carriers
which form the cable, and just as in the previous example, these contribute
enough to $E$ to make bound~(\ref{bound}) work (Sec.~\ref{coaxial}). 

Page revives\cite{page1} Deutsch's old objection to the entropy
bound to the effect that weak thermal excitation of the system
involves a violation\cite{Deutsch}.  I give in Sec.~\ref{low_temp} a brief
argument showing why the entropy bound is always obeyed by a {\it complete\/}
system, even at low temperature.  The problem  of entropy bound violations
at low temperatures is evidently a red herring !  The above argument
also illustrates how to deal with any mixed state which ascribes low
probabilities to the high energy pure states.
 
Page also revives the old argument\cite{UW,MFW,wald} that bound~(\ref{bound})
will be violated when there is a virtually unlimited number of
particle species in nature.  I have earlier refuted
this\cite{bek94,bek99,bek82,bek83}.   In addition, as I discuss in
Sec.~\ref{proliferation}, one knows today  that the vacuum of quantum field
theory is gravitationally unstable when many species of particles
exist\cite{BEF}.  Such instability makes Page's second revived argument moot.

\section{Nonlinear scalar field}
\label{field}

Page's first example\cite{page1,page2} deals with a self-interacting scalar
field with a multiwell potential, and considers configurations of the field
which vanish at a certain boundary of radius $R$.  For the symmetric double
well potential Page correctly points out that the classically degenerate
ground states of the field, each localized in one  well, engender, by quantum
tunnelling between the wells, a new ground state $\psi_0$ (energy $E_0$) with
equal amplitude at each well and a first excited state $\psi_1$ a very
small energy $E_1-E_0$ above the ground state.  He then hastily concludes
that this violates the bound because the entropy of a mixed state built out
of $\psi_0$ and $\psi_1$ can reach $\ln 2$ (ground and excited states equally
probable), while $(E_1-E_0)R$ can be very small in natural
units ($\hbar=c=1$).  In this interpretation Page considers the energy $E$
mentioned in the bound as the energy measured {\it above\/} the ground
state.  It would indeed be so if the ground state referred to a spatially
unrestricted configuration, because then the bottom of the potential well
would be the correct zero of energy (neglecting zero point fluctuations).  

But since the field is required to vanish at radius $R$, the energy $E_0$ of
the described ground state is a function of $R$, and it makes little sense to
take it as the zero of energy.  For example, by expanding the system can do
work ($-\partial E/\partial R\neq 0$), so that its gravitating energy changes,
and cannot be taken as zero for all $R$.  The gravitating energy for the
equally likely mixture of ground and excited states should be identified with
${\scriptstyle 1\over \scriptstyle 2}(E_0(R)+E_1(R))\approx E_0(R)$, which
does go to zero as $R\rightarrow \infty$, and is, therefore, properly
calibrated with respect to the global ground states of the theory.. The
exponential smallness of $E_1-E_0$, which Page pounced upon, is not very
relevant, as will transpire.

Since there are no solitons in $D=1+3$ spacetime, a finite sized
field configuration [the only interesting case---see (\ref{bound})] must be
confined by a ``wall''.  There are three parts to the energy
$E$ of the total system (before tunnelling is taken into account): the
classical energy $E_c$ of the field configuration concentrated around one
well but vanishing at radial coordinate $r=R$, the quantum correction
$E_v$ due to the zero point fluctuations about the classical configuration,
and $E_w$, the energy of the ``wall'' at $r=R$.  As I show below, $E_w$ is
at least of the same order as $E_c$, and both dominate Page's energy.

\subsection{Classical two-well configurations}
\label{two}

The double well potential field theory comes from the lagrangian
density
\begin{equation}  
{\cal L}=-{1\over 2}\partial_\mu\phi\,\partial^\mu\phi-{1\over
4}\lambda(\phi^2-\phi_m^2)^2
\label{theory}
.\end{equation}
This gives  the field equation
\begin{equation}  
\partial_\mu\,\partial^\mu\phi-\lambda\phi(\phi^2-\phi_m^2)=0
\label{TD}
.\end{equation}
Every spherically symmetric configuration inside a spherical
box of radius $R$ will thus satisfy (I use standard spherical coordinates;
$'$ denotes derivative w.r.t. to $r$)
\begin{equation}  
r^{-2}(r^2\phi')'-\lambda\phi(\phi^2-\phi_m^2)=0
\label{TI}
.\end{equation}
Regularity requires that $\phi'=0$ at $r=0$.  Page chooses $\phi=0$ at
$r=R$.  The energy of such a configuration will be
\begin{equation}  
E_c={\scriptstyle 1\over \scriptstyle 2}\int_0^R\left[\phi'^2+{\scriptstyle
1\over \scriptstyle 2}\lambda(\phi^2-\phi_m^2)^2\right] r^2\,dr
\label{Ec}
.\end{equation}

Since one is interested in the ground state, I require that $\phi$ have its
first zero at $r=R$.  Multiplying Eq.~(\ref{TI}) by $r^2\phi$ and integrating
over the box allows, after integration by parts and use of the boundary
conditions, to show that
\begin{equation}  
\int_0^R\phi'^2r^2\,dr=\lambda \int_0^R(\phi_m^2-\phi^2)\phi^2 r^2\,dr
\end{equation}
whereby
\begin{equation}  
E_c={\scriptstyle 1\over \scriptstyle
4}\lambda\int_0^R(\phi_m^4-\phi^4)r^2\,dr
\label{energy}
.\end{equation}

It proves convenient to adopt a new, dimensionless, coordinate
$x \equiv \surd\lambda\phi_m\,r$ and a dimensionless scalar $\Phi\equiv
\phi/\phi_m$.  Then Eq.~(\ref{TI}) turns into a parameter-less equation:
\begin{equation}  
{1\over x^2}{d\over dx}\left(x^2{d\Phi\over
dx}\right)+\Phi(1-\Phi^2)=0
\label{TI2}
.\end{equation}
Using $d\Phi/dx=0$ at $x=0$ one may integrate the equation to get
\begin{equation}  
{d\Phi\over dx}=-{1\over x^2}\int_0^x\Phi(1-\Phi^2)\,x^2\,dx.
\label{helpful}
\end{equation}
If the integration starts with $\Phi(0)>1$, then by
continuity the r.h.s. here is positive for small $x$, so that $\Phi$ grows. 
There is thus no way for the r.h.s. to switch sign, so $\Phi(x)$ is
monotonically increasing and can never have a zero.  If $\Phi(0)=1$, it is
obvious that the solution of Eq.~(\ref{helpful}) is
$\Phi(x)\equiv 1$ which cannot satisfy the boundary condition at $r=R$.  
Thus the classical ground state configuration we are after requires
$\Phi(0)<1$.

When $\Phi(0)<1$ it can also be seen from Eq.~(\ref{helpful}) that $\Phi$ is
monotonically decreasing with $x$.  For a particular $\Phi(0)$, $\Phi(x)$
will reach its first zero at a particular $x$ which I refer to as $x_0$. 
This can serve as the parameter singling out the solution in lieu of
$\Phi(0)$.  One thus has a family of ground state configurations
$\Phi(x,x_0)$.  Each such configuration corresponds to a box of radius
$R=x_0(\surd\lambda\phi_m)^{-1}$.  In terms of the new variables one can
write Eq.~(\ref{energy}) as
\begin{equation}
E_c ={x_0\over 4\lambda R}\int_0^{x_0}(1-\Phi^4)x^2\, dx.
\label{Ec2}
\end{equation}
The dependence  $E_c\propto \lambda^{-1}$ is well known from kink solutions
of (\ref{TD}) in $D=1+1$\cite{rajaraman}, where the role of $R^{-1}$ is
played by the effective mass of the field.   Numerical integration of
Eq.~(\ref{TI2}) shows that the factor $x_0\int_0^{x_0}(1-\Phi^4)x^2\, dx$
grows monotonically from 32.47 for $\Phi(0)= 0\ (x_0=3.1416)$ to 232.23 for
$\Phi(0)=0.98\ (x_0=5.45)$ to infinity as $\Phi(0)\rightarrow 1\
(x_0\rightarrow\infty)$. Since $E_c$ is not exponentially small, the quantum
tunnelling corrections that Page discussed are negligible, so one need only 
add to $E_c$ the zero point fluctuations energy $E_v$ plus the wall energy
$E_w$ to get the full energy of the ground state, $E_0$.  This plays the
role of $E$ in the bound (\ref{bound}). 

I shall not bother to calculate $E_v$.  This can be done by present
techniques only for the weak coupling case $\lambda<1$\cite{rajaraman}.  It
is then found in other circumstances, e.g. the $D=1+1$ kink, that
$E_v$ is small compared to $E_c$.  The situation for large $\lambda$ (the
strong coupling regime) is unclear.  However, it is appropriate to recall
here that the theory (\ref{theory}) is trivial in that it makes true
mathematical sense only in the case $\lambda=0$\cite{callaway}.  Theorists
use it for $\lambda\neq 0$ to obtain insights which are probably trustworthy
in the small $\lambda$ regime, but probably not for large $\lambda$.

I now set a lower bound on $E_w$.  A look at Eq.~(\ref{Ec2}) shows that for
$\Phi(0)\ll 1$ and so $\Phi(x)\ll 1 $), $E_c$ scales as $x_0^4/R\propto
R^3$. Numerically the exponent here only drops a little as $\Phi(0)$
increases; for example, it is 2.86 for $\Phi(0)=0.98$.  So I take it as 3. 
On virtual work grounds (consider expanding $R$ a little bit) the $R^3$
dependence means the $\phi$ field exerts a suction (negative pressure) of
dimension $\approx (3E_c/4\pi R^3)$ on the inner side of the wall.  By
examining the force balance on a small cap of the wall, one sees that in
order for the wall to withstand the negative pressure, it must support a
compression (force per unit length) $\tau\approx (3E_c/8\pi
R^2)$\cite{bek82}.  Under this compression vibrations on the wall will
propagate superluminally unless the surface energy density is at least as big
as $\tau$ (dominant energy condition).  Thus one may conclude that the wall
(area $4\pi R^2$) must have (positive) energy $E_w> 3E_c/2$ which {\it
adds\/} to $E_c$ to give
$E>5E_c/2$.   

For $\Phi(0)$ very close to unity the coefficient is somewhat lower than
$5/2$.  However, by then $E_c R$ is already much larger than the
corresponding quantity for $\Phi(0)\ll 1$  (six times larger for
$\Phi(0)=0.98$). Using the value of $E_c$ for $\Phi(0)\ll 1$ from
the preceding argument, I thus conclude that for all physically relevant 
$\Phi(0)$, $2\pi ER > 127.5\lambda^{-1}$.  This is certainly not exponentially
small with $R$ as Page claimed !  True, formally  it seems possible to have a
violation of the bound on the entropy $\ln 2$ of the 50\% mixture of
ground and excited states whenever $\lambda> 127.5/\ln 2 =183.95$.  However,
this is the strong coupling regime.  For all one knows the zero point energy
may then become important and tip the scales in favor of the entropy bound. 
At any rate, because the theory (\ref{theory}) is trivial, one is more likely
to be overstepping here the bounds of its applicability than to be witnessing
a violation of the entropy bound at large $\lambda$.

In summary, we have found that whenever the calculation is meaningful
($\lambda$ not large), the entropy bound (\ref{bound}) is satisfied
provided $E$ includes all contributions to the energy.  In his
later defense against this observation, Page\cite{page2} cites my paper
with Schiffer\cite{SB} as an excuse for including in $E$ just the excitation
energy above the classical ground state.  However, he neglects to point out
that we ourselves restricted use of this approach to an assembly of quanta
of a massless {\it noninteracting\/} field.

I should also mention that in $D=1+1$ spacetime it is possible to find
analytically all static classical configurations (and their energies) for the
theory (\ref{theory}) in a box, and the distribution of energy levels
is such that the entropy bound is sustained\cite{bek_guen}. 

\subsection{Multiwell potential}
\label{multi}

Page also confronts bound (\ref{bound}) with a theory like 
(\ref{theory}) but with a potential having three equivalent
wells. Pressumably one would like one of these centered at $\phi=0$, with the
other two flanking it symmetrically.  Then Page's conclusion that there are
three exponentially close states (in energy) is untenable.  This would
require three classically degenerate configurations, which certainly exist in
open space (field $\phi$ fixed at one of three well bottoms).  However, one
is here considering a finite region of radius $R$ with $\phi=0$ on the
boundary.  One exact solution is indeed $\phi=0$, and it has zero energy (the
zero point fluctuation energy correction will, however, depend on $R$).  Then
there are two degenerate solutions in which the field starts at $r=0$ in one
side well and then moves to the central one with $\phi\rightarrow 0$ as
$r\rightarrow R$.  By analogy with our earlier calculations, the common
classical energy of these two configurations will be of  $O\left(8(\lambda
R)^{-1}\right)$.  It cannot thus be regarded as the zero of energy; this role
falls to the energy of the $\phi=0$ configuration.  When tunnelling between
wells is taken into account, one has a truly unique ground state and two
excited states of classical origin split slightly in energy (plus the usual
gamut of quantum excitations).  The entropy of an equally weighted mixture of
these states is $\ln 3$.  The mean energy $E$ is ${\scriptstyle 2\over
\scriptstyle 3}(E_c+E_w)$ of an excited state, that is $E=O(10(\lambda R)^{-1})$, so the
entropy bound is satisfied, at least in the weak coupling regime where the
theory makes sense.

When the potential has $n=5,7,9,\cdots\ $ equivalent wells with one
centered at $\phi=0$ and the rest disposed symmetrically about it, there
will be a single zero-energy configuration ($\phi\equiv 0$), and
${\scriptstyle 1\over \scriptstyle 2}(n-1)$ pairs of degenerate configurations
with succesively ascending $R$-dependent energies.  For $n=4,6,8,\cdots\ $
wells there is no zero-energy configuration, but there are
${\scriptstyle 1\over \scriptstyle 2}n$ pairs of degenerate configurations
with $R$ dependent energies.  Because of the extra energy splitting appearing
here already classically, I expect, by analogy with the previous results, 
that the appropriate mean configuration energy (perhaps supplemented by wall
energy), when multiplied by $2\pi R$, will bound the $\ln n$ entropy from
above. 

\section{Electromagnetic Onion}
\label{onion}

As a second counterexample to the entropy bound, Page proposes\cite{page1} a
sphere of radius $R$ partitioned into $n$ concentric shells; the partitions
and the inner and outer boundaries are regarded as infinitely conducting. 
He points out that the lowest ($\ell=1$) three magnetic-type
electromagnetic modes in the shell of median radius $r$ have frequency
$\omega\approx 1/r$.  Since there are $3n$ such modes (three for each
shell), Page imagines populating now one, then another and so on with a
single photon of energy
$\sim 1/r$ for the appropriate $r$.  These one-photon states allow him to
form a density matrix which, for equally weighted states, gives entropy
$\ln (3n)$ and mean energy $\sim 2/R$ (since $R/2$ is the median radius of
the shells if they are uniformly thick).  Page concludes that bound (1) is
violated because the entropy grows with $n$ but the mean energy does not. 

Page has again missed out part of the energy.  The modes he needs owe their
existence to the infinitely conducting partitions that confine them,
each to its own shell.  Were passage between shells possible, then old
work\cite{bek81} already established that the entropy bound works for the
electromagnetic field  confined to an empty sphere (or for that matter
confined to any parallelepiped\cite{bek84}. To be highly conducting, the
envisaged partitions must contain a certain number of charge carriers, whose
aggregated masses turn out to contribute enough to the  system's total energy
$E$ to make it as large as required by the entropy bound (\ref{bound}). 
Ignoring the masses of the charge carriers goes against the condition that
the bound applies to a complete system: the carriers are an essential
component, so their gravitating energy has to included in $E$.

I assume all partitions to have equal thickness $d$.  One mechanism that
can block the waves from crossing a partition is a high plasma frequency
$\omega_p$ of the charge carriers in the partitions.  We know\cite{jackson}
that in a plasma model of a conductor with collisionless charge carriers, the
electromagnetic wave vector for frequency $\omega$ is
$k=\omega(1-\omega_p^2/\omega^2)^{1/2}$, so that if $\omega<\omega_p$, the
fields do not propagate as a wave.  Nevertheless they do penetrate a distance
$\delta=\omega^{-1}(\omega_p^2/\omega^2-1)^{-1/2}>\omega_p{}^{-1}$ into the
plasma before their amplitudes become insignificant.  In order to prevent
these evanescent waves from bridging a partition, one must thus require
$\delta<d$, i.e., $\omega_p d>1$.  But 
\begin{equation}
\omega_p{}^2=4\pi{\cal N}e^2/m
\label{plasma}
\end{equation} where ${\cal N}$ is the density of charge
carriers of charge $e$ and mass $m$.  Since $d<R/n$, all this gives us
$(4\pi R^2 d){\cal N}>m n^2 d/e^2$.  Now $4\pi R^2 d$ is the volume of
material in the outermost partition.  Properly accounting for the variation
of partition area with its order $i$ in the sequence (we employ the sum $\sum
i^2$),  tells us that for $n\gg 1$ the total mass-energy in charge carriers
in all the partitions is $E\approx nm(4\pi R^2 d/3){\cal N}$.  Substituting
our previous bound on $(4\pi R^2 d){\cal N}$, I get $E>{\scriptstyle 1\over
\scriptstyle 3}n^3 m^2 d/e^2$.  

Now a charge carrier's Compton length has to be smaller than $d$, for
otherwise the carriers would not be confined to the partitions; hence
$md>1$.  As a matter of principle $e^2<1$, because more strongly coupled
electrodynamics makes structures, e.g. atoms and partitions, which are held
together electrically, unstable\cite{rozental} (in our world $e^2<10^{-2}$). 
Therefore, since $d<R/n$ one gets $E>{\scriptstyle 1\over \scriptstyle
3}n^4R^{-1}$, which strongly dominates $\sim 2/R$, the energy in photons
that Page gets.  In particular $2\pi RE>2 n^4$, which is always much larger
than the  entropy in photons $\ln (3n)$. 

The only alternative mechanism for keeping electromagnetic waves from
penetrating into a conductor is the skin effect\cite{jackson}.  The skin
depth is $\delta\approx(2\pi\omega\sigma)^{-1/2}$, where $\sigma$ is the
conductivity.  In the simple Drude model\cite{jackson}, $\sigma={\cal
N}e^2(m/\tau-\imath m\omega)^{-1}$, where $\tau$ is the slowing-down
timescale for a charge carrier due to collisions, and $\imath=\sqrt{-1}$.  The
formula for $\delta$ refers to an Ohmic (real) conductivity rather than an
inductive (imaginary) one.  Thus one must demand that $\omega\ll 1/\tau$. 
But then 
\begin{equation}
\delta\gg (2\pi{\cal N}e^2/m)^{-1/2}.
\label{skin}
\end{equation}
As before one must require $\delta<d<R/n$.  This gives  $(4\pi R^2 d){\cal
N}\gg{\scriptstyle 1\over \scriptstyle 2}m n^2 d/e^2$ which is just a
stronger version of the lower bound on ${\cal N}$ we got before.   Repeating
the previous discussion {\it verbatim\/} shows that $2\pi RE\gg n^4$, which
bounds Page's $\ln (3n)$ entropy confortably.

In conclusion, bound (\ref{bound}) is satisfied by the system photons $+$
charge carriers.  It should be clear that the entropy of the conducting
material with its many carriers, which we have been ignoring, may well
dominate that in photons.  Here one can fall back on the usual
argument\cite{bek81} that in a random assembly of particles the entropy is
of order of the number of particles, and that each particle's Compton length
is necessarily smaller than the system's radius.  From these two conditions
it follows that the entropy bound applies---with room to spare---to the
charge carrier system by itself.  Putting all this together makes it
clear that it applies to the complete onion system as well.

\section{Coaxial cable loop}
\label{coaxial}

Page's third example\cite{page2} is furnished by an electromagnetic
field confined to a coaxial cable of length $L$ which is coiled up so as to
fit within a sphere of radius $R$, with $R\ll L$, before being connected end
to end to form a closed loop.  Page's entirely qualitative reasoning
proceeds by analogy with a rectilinear coaxial cable with periodic boundary
conditions. A rectilinear infinitely long coaxial cable has
some electromagnetic modes which propagate along its axis with arbitrarily low
frequency.  Page notes that for the coiled-up cable, each right moving
mode is accompanied by a degenerate (in frequency) left
moving mode (basically this follows from time reversal invariance of
Maxwell's equations).  He then argues that if the cable's {\it outer\/} radius
$\varrho$ is thin on scale $R$, the structure of the electromagnetic modes is
little affected by the cable's curvature. This leads him to estimate the
lowest frequency $\omega_1$ as similar to that of the rectilinear coaxial
cable with periodic boundary conditions with period $L$: $\omega_1\approx
2\pi L^{-1}$.  Page then imagines a mixed electromagnetic state of energy
$E-E_{\rm vac}=2\pi L^{-1}$ in which a single photon occupies the right- or
the left-moving mode of frequency
$\omega_1$ with equal probabilities ${\scriptstyle 1\over\scriptstyle 2}$. 
The entropy of this state is $\ln 2$.  However, $2\pi(E-E_{\rm vac})R\approx
4\pi^2 (R/L)$ which could be very small compared to $\ln 2$.  Therefore,
Page exhibits this example as a violation of the Schiffer-Bekenstein `strong
entropy bound'.

As I stressed in Sec.\ref{intro}, the strong bound was never formalized in a
theorem for electromagnetism.  If Page is correct in his estimate  of
$\omega_1$, such a theorem cannot apply to a cavity with not simply connected
crossection, like the coaxial cavity.  (However, the myriad examples studied
numerically\cite{bek84,review} strongly suggest that a theorem of the
desired type should exist for simply connected cavities.)  At any rate, in
the present example the interesting question is whether the coaxial cavity
plus electromagnetic field obeys the original entropy bound (\ref{bound}).

The inner conductor of the cable---let its outer radius be $\rho$ and its
thickness $d$---is an essential part of the system, for without it the lowest
propagating frequency would be $O(\varrho^{-1})$, i.e., very large on scale
$L^{-1}$.  Now in order for the inner conductor to keep the fields out of it,
as required by the whole notion of a coaxial cable, one must have either
$\rho>d>\omega_p^{-1}$ or $\rho>d>\delta$ (see Sec.~\ref{onion}).  

In the case $\rho>d>\omega_p^{-1}$, Eq.~(\ref{plasma}) informs us that ${\cal
N}\rho d>m(4\pi e^2)^{-1}$ where, as in Sec.~\ref{onion}, $m$ denotes a
charge carrier's mass and ${\cal N}$ the carrier density.  Since the volume
of material in the inner conductor is $\pi L[\rho^2-(\rho-d)^2]$, its mass
energy is at least $\pi m L{\cal N}\rho d$, and thus the total mass-energy
$E$ of cable plus field is  constrained by $E>m^2 L(4 e^2)^{-1}$.  But a 
charge carrier has to
be localized within the conductor, which requires that $md>m\rho\gg 1$. 
Hence $E\gg L(4 e^2 \rho^2)^{-1}$ so that $2\pi ER\gg
(\pi/2e^2)(L/\rho)(R/\rho)$.  Now obviously $R>\varrho>\rho$ and $L\gg R$ by
the conditions of the problem, while $e^2<1$ by the condition of stability
(see Sec.~\ref{onion}), so that $2\pi ER\gg 1$.  Thus the bound on entropy
confortably bounds Page's entropy $\ln 2$.  

Were the mixed state instead to involve one photon in any one of the low lying
modes having no crossectional nodes and wavelength along the cable of the form
$k (L/2\pi)$ with $k$ an integer, one could get a bigger entropy.  There are
$O(L/\rho)$ such modes with frequency below that of the lowest lying
transversally excited mode (which is obviously of order
$\rho^{-1}$), so the entropy of the envisaged state is
$\sim\ln(L/\rho)$.  Because $\ln x< x$ for $x> 1$, this is
obviously bounded by $(L/\rho)(R/\rho)$ and, therefore, by $2\pi ER$. 

In the case $\rho>d>\delta$, Eq.~(\ref{skin}) gives ${\cal N}\rho d\gg m(2\pi
e^2)^{-1}$.  This is just a stronger version of the earlier lower
bound on ${\cal N}$.   Repeating the previous discussion shows again that
$2\pi ER$, with $E$ the total energy of the system, bounds the entropy. 

\section{Low temperature systems}
\label{low_temp}

Page also examined\cite{page1} the generic example in which the system's
density matrix is diagonal and involves $g+1$ equally probable pure states. 
The entropy is $\ln(g+1)$.  Obviously the greatest challenge to bound (1)
is posed when $g$ of the states are degenerate and just a small energy
$\Delta$ above the (unique) ground state whose energy is $\epsilon_0$.  Now
for any $\Delta$ and $g$ the mean energy $E$ of the complete system is {\it
at least\/} $\epsilon_0$, so $2\pi RE>2\pi R\epsilon_0$.  The
present system is so generic that it is not feasible to estimate
$\epsilon_0$ as I did previously. It should be clear, however, that the
system's longest possible Compton wavelength, $\epsilon_0^{-1}$, should lie
well below $R$; otherwise the contention that the system fits within a
definite radius $R$ would make no sense as it would be poorly localized on
scale $R$.  It is probably conservative to take $R\epsilon_0>3$. 
Therefore,  no violation of the entropy bound can occur for $g<10^8$.  Can
$g$ be bigger ?   

Now quantum mechanical systems have low degeneracies in the
low lying levels.  Quantum field systems have more.  But even in those
systems with accidental degeneracies, e.g. electromagnetic field in a
cubical box, $g<10$\cite{bek84}.  As the end of Sec.\ref{multi} illustrates,
it is hard to generate high degeneracies in nonlinear fields in a box.  One
could create a highly degenerate state artificially by lumping together
states closely spaced in energy, e.g. the one-photon states using the modes
with $k=1,2, \cdots$ of the coaxial cable. But as shown in Sec.\ref{coaxial},
the large $L/\rho$ required to have many of these modes defeats the attempt
to keep $2\pi RE$ small because a long cable involves a lot of charge
carriers.  This illustrates the point that, unlike Page, one cannot
arbitrarily legislate a high degeneracy. Rather, one must examine the
energy cost paid by the passive components of the system in providing a large
number of closely spaced states for the active part which can be consolidated
into a formally highly degenerate one.     

In a related line of thought, Page\cite{page1} revives an old challenge to
the entropy bound\cite{Deutsch}, which is occasionally discovered
anew\cite{MFW,wald}: a  system in a thermal state seems to violate the
entropy bound if its {\it inverse\/} temperature $\beta$ is sufficiently
high.  The essence and resolution of the problem is captured by the
following purely analytical treatment.  Consider a system of radius $R$ with
ground energy $\epsilon_0$, a $g$-fold degenerate excited state at energy
$\epsilon_1=\epsilon_0+\Delta$, and higher energy states.  For sufficiently
large $\beta$ one may neglect the higher energy states in the partition
function $Z=\sum_i\exp(-\beta\epsilon_i)$, and so approximate it by
$\ln Z\approx -\beta\epsilon_0 + \ln(1+g e^{-\beta\Delta})$.  The {\it
mean\/}  energy is
\begin{equation}
E=-{\partial\ln Z\over \partial\beta}=\epsilon_0+{g\Delta\over
e^{\beta\Delta}+g}
\end{equation}
while the entropy takes the form 
\begin{equation}
S=\beta E+\ln Z={g\beta\Delta\over e^{\beta\Delta}+g}+\ln(1+g
e^{-\beta\Delta})
.\end{equation}
The typical claim\cite{page1,Deutsch} is``measure energies from the ground
state so that $\epsilon_0=0$; then for $\beta>2\pi R$ one gets $S>2\pi RE$
and the bound is violated''. 

But we have already seen in Sec.~\ref{field} that taking the zero of energy
of a system at its ground state is not automatically justified because it may
mean using as $E$ something distinct from the gravitating energy.  And I
have already discussed why $R\epsilon_0$ should be at least of the order a
few, say 3.   

The interesting quantity now is 
\begin{equation}
S-2\pi RE=\Xi(\beta\Delta)\equiv{g(\beta\Delta-2\pi R\Delta)\over
e^{\beta\Delta}+g}+\ln(1+g e^{-\beta\Delta})-2\pi R\epsilon_0
.\end{equation}
The function $\Xi(y)$ has a single {\it maximum\/} at $y=2\pi R\Delta$ where
$\Xi=
\ln(1+g e^{-2\pi R\Delta})$. I thus conclude that
\begin{equation}
S<2\pi RE+[\ln(1+g e^{-2\pi R\Delta})-2\pi R\epsilon_0].
\label{last}
\end{equation}
For the quantity in square brackets to be nonnegative it would be necessary
for $g\geq e^{2\pi R\Delta}[e^{2\pi R\epsilon_0}-1]$, i.e., $g>10^8$. As we
saw above, this cannot be arranged.  Thus the quantity in
square brackets in Eq.~(\ref{last}) has to be negative: {\it for
sufficiently low temperature the entropy bound is upheld\/} with room to
spare.  The above argument also illustrates how to deal with any mixed state
which ascribes low probabilities to the high energy pure states.

Early realistic numerical calculations of thermal quantum fields in
boxes\cite{bek83} did reveal that,  were the ground state energy to be
ignored, the bound on entropy would be violated at very low temperatures,
typically when  $E<10^{-9}R^{-1}$ ($R$ enters through the ``energy gap''
$\Delta$).  It was also clear early\cite{bek81,bek83} that taking any
reasonable ground state energy into account precludes the violation.  As the
temperature rises, more and more pure states are excited, and eventually
$S/E$ peaks and begins to decrease.  In this regime the entropy bound is
always obeyed regardless of whether or not one includes $\epsilon_0$
in the total energy\cite{bek84}.
 
\section{Proliferation of species}
\label{proliferation}

Page also revives the old ``proliferation of species'' challenge to
the entropy bound\cite{UW,MFW}.  Suppose there were to exist as many copies
$N$ of a field e.g. the electromagnetic one, as one ordered.  It seems as if
the entropy in a box containing a fixed energy allocated to the said fields
should grow with $N$ because the bigger $N$ is, the more ways there are to
split up the energy.  Thus eventually the entropy should surpass the entropy
bound.  Numerical estimates show that it would take $N\sim 10^9$ to do the
trick\cite{bek83}. A similar picture seems to come from Eq.~(\ref{last});
the degeneracy factor $g$ should scale proportionally to $N$ making the
factor in square brackets large, so that, it would seem, one could not use
the argument based on (\ref{last}) to establish that $S<2\pi RE$. However,
as recognized in refs. \cite{bek94,bek99,bek82}, the above
reasoning fails to take into account that each field species makes a
contribution of zero point fluctuations energy which gets lumped in
$\epsilon_0$.  If these contributions are positive, then the negative term
in the square bracket eventually dominates the logarithm as $N$ grows, and
for large $N$ one again recovers the entropy bound\cite{bek94}.  If they are
negative (which implies a Casimir suction proportional to $N$ on the walls
which delineate the system), then the scalar field example suggests that the
wall energy, which must properly be included in $\epsilon_0$, should
suffice to make the overall $\epsilon_0$ positive\cite{bek94}.  Again the
entropy bound seems safe.

There is an alternative view\cite{bek82-2,bek99}: the seeming clash
between entropy bound and a large number of species merely tells us that
physics is consistent only in a world with a limited number of species, such
as the one that is observed.  Indeed, as Brustein, Eichler and
Foffa\cite{BEF} have argued, a large number of field species will make the
vacuum of quantum field theory unstable against collapse into a ``black hole
slush''.  Thus the proliferation of species argument against the entropy
bound is not even physically consistent.

\section{Caveats}

I have stressed the robustness of the entropy bound for weakly gravitating
systems.  But one should recall that the  bound has its limitations.  These
principally belong to the strongly gravitating system regime.  Bound
(\ref{bound}) does not apply in wildly dynamic situations such as those
found inside black holes\cite{bousso}, and it is not guaranteed to work for
large pieces of the universe (which, after all, are not complete systems).

\acknowledgments
This work was supported by the Hebrew University's
Intramural Research Fund.  I thank Eduardo Guendelman  for much information
and Mordehai Milgrom for incisive critical remarks.

\end{document}